\documentstyle[12pt]{article} 
\begin{document}
\pagestyle{plain}
\setcounter{page}{1}

\newcommand{\be}{\begin{equation}}
\newcommand{\ee}{\end{equation}}
\newcommand{\bea}{\begin{eqnarray}}
\newcommand{\eea}{\end{eqnarray}}
\renewcommand{\theequation}{\arabic{section}.\arabic{equation}}
\newcommand{\nono}{\nonumber}
\newcommand{\cF}{{\cal F}}
\newcommand{\tcF}{\tilde{{\cal F}}}
\newcommand{\tF}{\tilde{F}}
\newcommand{\tf}{\tilde{f}}
\newcommand{\tB}{\tilde{B}}
\newcommand{\ta}{\tilde{a}}
\newcommand{\tg}{\tilde{g}}
\newcommand{\bG}{\bar{G}}
\newcommand{\bB}{\bar{B}}
\newcommand{\bq}{\bar{q}}
\newcommand{\ba}{\bar{a}}
\newcommand{\si}{\sigma}
\newcommand{\del}{\partial}
\newcommand{\jac}[2]{\frac{\partial #1}{\partial #2}}
\newcommand{\str}{2\pi\alpha'}
\newcommand{\dJ}{\delta J}
\newcommand{\hA}{\hat{A}}
\newcommand{\hlam}{\hat{\lambda}}
\newcommand{\Tr}{{\rm Tr}}
\newcommand{\tr}{{\rm tr}}
\newcommand{\ket}[1]{|#1\rangle}
\newcommand{\sta}[1]{|#1)}
\newcommand{\dDst}{\int_{\Sigma}d^2\si}
\newcommand{\PB}[2]{\{#1,#2\}_{Y^{-1}}}
\newcommand{\Yi}{Y^{-1}}

\setcounter{section}{0}
\setcounter{subsection}{0}
\def\thefootnote{\fnsymbol{footnote}}
\begin{titlepage}


\begin{flushright}
KEK Preprint 99-160 \\
January 2000 \\
\tt hep-th/0001011
\end{flushright}
\vspace{1cm}
\begin{center}
\huge
Noncommutativities of D-branes \\
and $\theta$-changing Degrees of Freedom in D-brane Matrix Models
\end{center}
\vspace{1cm}
\normalsize
\begin{center}
{\sc Tsunehide Kuroki}
\footnote{
e-mail address:\ \  {\tt kuroki@tanashi.kek.jp}}\\
\vspace{0.3cm}
{\it Institute of Particle and Nuclear Studies \\
High Energy Accelerator Research Organization (KEK), \\
Tanashi branch, Tokyo 188-8501, Japan}
\end{center}
\vspace{7mm}
\begin{center}
{\large Abstract}
\end{center}
\noindent
It is known that when there are several D-branes, their space-time 
coordinates in general become noncommutative. From the point of view of 
noncommutative geometry, it reflects noncommutativity of the world volume 
of the D-branes. On the other hand, as we showed in the previous work, 
in the presence of the constant antisymmetric tensor field 
the momentum operators of the D-branes have noncommutative structure. 
In the present paper, we investigate a relation 
between these noncommutativities and the description of D-branes 
in terms of the noncommutative Yang-Mills theory recently proposed 
by Seiberg and Witten. It is shown that the noncommutativity 
of the Yang-Mills theory, which implies that of the world volume 
coordinates, originates from both noncommutativities of the transverse 
coordinates and momenta from the viewpoint of the lower-dimensional D-branes. 
Moreover, we show that this noncommutativity is transformed 
by coordinate transformations on the world volume and thereby can be chosen 
in an arbitrary fixed value. 
We also make a brief comment on a relation between this fact and 
a hidden symmetry of the IIB matrix models. 
\vspace{7mm}
\end{titlepage}


\newpage
\section{Introduction}
Candidates for the nonperturbative definition of string theory have been 
proposed for these few years\cite{BFSS,IKKT,Mald}. Although these matrix 
models have passed several nontrivial checks so far, they are far from 
satisfactory because not only they have little predictions 
as to nonperturbative effects of string theory, but they do not completely 
reproduce the perturbative string theory. These failure might result 
from the lack of information on fundamental degrees of freedom 
and symmetry or principle which governs them in these models. 
Since matrix models start with the lower-dimensional D-branes 
such as D-instantons or D-particles as fundamental degrees of freedom
(although there is a difference in their interpretations), 
it is important to examine the physics of D-branes and survey true fundamental 
degrees of freedom and a hidden symmetry in the nonperturbative string theory. 
   
One of the most appealing features of D-branes is their noncommutative 
structure. Namely, when there are several D-branes which are parallel 
to each other, their transverse coordinates are promoted to non-commuting 
matrices\cite{boundstate}. {}From the point of view of noncommutative 
geometry, this noncommutativity can be considered to reflect 
the noncommutative structure of the world volume of D-branes,  
since the transverse coordinates are functions on the world volume 
of D-branes which is naturally described by them.  In what follows, 
we will refer to this noncommutativity as {\it transverse noncommutativity}.

On the other hand, we found in our previous work\cite{KK} 
that in the presence of the background antisymmetric tensor field 
along the transverse directions the momentum operators of the D-brane 
toward these directions become noncommutative.\footnote{
Accordingly Matrix theory or a low energy effective field theory of 
D-branes on a torus in the constant antisymmetric field background 
is described by a gauge theory on a noncommutative torus\cite  
{CDS,DH}.} 
Henceforth this noncommutativity will be referred to as 
{\it momentum noncommutativity}. 
These two noncommutativities are dual to each other 
in the sense that the former realizes the noncommutative structure 
of the transverse coordinates of world volume, while the latter realizes 
that of the conjugate momenta. Moreover, as shown in \cite{KK}, when the 
transverse directions are compactified on the two-torus they satisfy 
non-trivial relations. These results are in nonperturbative aspects 
of string theory and should be naturally considered to reflect 
a hidden symmetry or an unknown mechanism of string theory. 
In fact, from the viewpoint of the underlying principle of string theory, 
there exist a proposal that regardless of fundamental strings or D-branes, 
their transverse directions and longitudinal ones have uncertainty 
of order of string scale and that play dual roles to each other, 
which is a manifestation of the fundamental principle or symmetry 
of string theory\cite{stur,yoneya,jeviyone}. 
It is natural to guess that noncommutativities of D-branes are closely 
related to this idea. Therefore further studies of these noncommutativities 
would shed light on future investigations. 

Recently Seiberg and Witten analyzed the open string theory in the presence 
of the constant Neveu-Schwarz 2-form field $B_{ij}$ and showed that 
its low-energy effective theory is described by the ordinary Yang-Mills (YM) 
theory when one adopts the Pauli-Villars regularization, while it is 
described by the noncommutative Yang-Mills (NCYM) theory with a 
parameter of noncommutativity given by $\theta=B^{-1}$ in the case of 
point-splitting regularization and thus both are physically equivalent
\cite{SW}. Here the NCYM theory with $\theta$ is defined by replacing 
all the ordinary product of functions such as gauge fields 
with what is called the $*$-product
\be
f(x)*g(x)=e^{\frac{i}{2}\theta^{ij}\jac{~}{\xi^i}\jac{~}{\zeta^j}}
          f(x+\xi)g(x+\zeta)|_{\xi=\zeta=0}, 
\label{*product}
\ee
where $\theta^{ij}$ is a real constant number. Noncommutative gauge symmetry 
with gauge parameter $\hlam$ for the noncommutative gauge field $\hA_i$ 
is given as 
\be
\hat{\delta}_{\hlam}\hA_i=\del_i\hlam+i\hlam*\hA_i-i\hA_i*\hlam.
\label{nctransf}
\ee
These implies the following non-trivial commutation relation 
of the base space coordinate $x^i$: 
\be
[x^i, x^j]=i\theta^{ij}.
\label{qplane}
\ee
In \cite{SW}, based on these observations, a concrete map 
from ordinary gauge fields to noncommutative ones is constructed.   
Furthermore, it is conjectured that there exists a series of equivalent NCYM 
theories with arbitrary values of $\theta$ and an interesting prediction 
is made that such a NCYM theory with $\theta$ between $\theta=0$ (ordinary) 
and $\theta=B^{-1}$ corresponds to a suitable regularization 
which interpolates between Pauli-Villars and point-splitting. 
In the following, we will call the noncommutativity $\theta$ 
of NCYM theory {\it longitudinal noncommutativity} because 
it is caused by the noncommutativity of the coordinates of the base space 
as in (\ref{qplane}) which are nothing but the longitudinal coordinates 
of the D-brane world volume as shown explicitly in \cite{SW,iran,chuho,DQ}. 

{}For an attempt to define string theory nonperturbatively 
from lower-dimensional D-branes as a fundamental constituent like
\cite{IKKT,yoneya,nb,topmm}, it would be meaningful to reinterpret 
the above results in terms of them. In fact, it is shown that 
higher-dimensional D-branes can be regarded as a configuration of 
infinitely many lower-dimensional ones with the transverse noncommutativity
\cite{ishi,ishi2,dhn,townsend}. 
In section 2, a D-string which is made from infinitely many D-instantons 
in the presence of a constant $B_{ij}$ background are considered 
as a simple example, which has been analyzed in our previous paper\cite{KK}. 
Then We apply the result in \cite{SW} to this configuration 
and interpret it in terms of D-instanton degrees of freedom. 
In particular, we clarify the interrelation 
between the above-mentioned noncommutativities. 
Similar analysis was also done in a recent paper\cite{IIKK}. 
In section 3, we focus on the degrees of freedom to change the value 
of $\theta$ suggested in \cite{SW} and refine it from the standpoint 
of the longitudinal noncommutativity. Section 4 is devoted to 
discussions on the meaning of our results for a hidden symmetry 
of matrix models. 

\section{Interrelation between Noncommutativities} 
\setcounter{equation}{0}

In this section we consider a configuration of infinitely many D-instantons 
in flat space with a metric $G'_{ij}$ in the presence of a constant 2-form 
field $B'_{ij}$  background. For simplicity, the coordinates of D-instanton 
$q^i$ ($\infty\times\infty$ matrices) are assumed to satisfy 
\be
[q^1, q^2]=-ik,
\label{wvnc}
\ee
where $k$ is a real number. In this case, since the components of $B'_{ij}$ 
not along the 1,2-directions can be gauge away, we can assume that 
$B'_{12}=-B'_{21}\neq 0$ and $B'_{ij}$=0 otherwise. {}From now on 
indices $i,j$ are understood to be 1 or 2. $k$ is nothing but a parameter 
which measures the transverse noncommutativity of D-instantons. 
On the other hand, $B'_{ij}$ parametrizes the momentum noncommutativity 
since the momentum operator of D-instantons $\pi_i$ satisfies 
\be
[\pi_1, \pi_2]=iB'_{12},
\label{momnc}
\ee
when $k=0$\cite{KK}.\footnote{
In \cite{KK} D-instantons are compactified on the $T^2$ with the $B_{ij}$ 
flux and (\ref{momnc}) is obtained reflecting the global structure of the 
torus. Thus it is likely that the existence of components of $B'_{ij}$ 
along compact directions which cannot be gauged away is essential 
to define the  momentum noncommutativity in a well-defined manner.}
When $k\neq 0$, (\ref{momnc}) are modified as 
\be
[\pi_1, \pi_2]=i\frac{B'_{12}}{1-kB'_{12}}. 
\ee
However, even if $k=0$, $\pi_i$ still has the noncommutativity 
by the presence of $B'_{12}$. Therefore we also regard it as a parameter 
of the momentum noncommutativity in the case of non-zero $k$. 

As shown in \cite{KK,IIKK}, as a background of string, our configuration 
of D-instantons is equivalent to a D-string in the background 
$g_{ij}$, $B_{ij}$, $F_{ij}$ given by\footnote{
Although in \cite{KK} the configuration is compactified on $T^2$ 
as mentioned in the above footnote, a similar argument also works 
in this case. }
\be
g_{ij}=G'_{ij},~~~~~
B_{ij}+F_{ij}=B'_{ij}+\frac{1}{k}\epsilon_{ij}.\label{D-1andD1}\footnote{ 
In order to match the notation with those in \cite{SW}, 
we change the normalizations of $B'_{ij}$, $B_{ij}$ from those in \cite{KK} 
by a factor $2\pi\alpha'$.}
\ee
In these equations, in the D-instanton picture in the right hand side 
a noncommutative world volume spanned by $q^1$, $q^2$ is constructed, while 
in the left hand side a D-string world volume in the ordinary static gauge 
is considered. In fact, if we set 
\be
B_{ij}=B'_{ij},~~~F_{ij}=\frac{1}{k}\epsilon_{ij}, 
\label{spiden}
\ee 
by doing the appropriate gauge fixing, we find that the D-instanton picture
corresponds to the choice of the coordinate (parametrization) 
on the D-string world volume in which $U(1)$ gauge field strength $F_{ij}$ 
is always equal to $\epsilon_{ij}/k$ \cite{ishi,ishi2,cornalba}. 
As we will see in the next section, if we choose a world volume coordinate
using the general coordinate transformation on the world volume 
in such a way that $F_{ij}$ is a constant, it turns out to have 
noncommutative structure in general. 
Therefore, since the world volume in the D-instanton picture is 
noncommutative, we should apply the result in \cite{SW} 
to the left hand side of (\ref{D-1andD1}), namely, a D-string in the 
static gauge. 

Seiberg and Witten began with the open string theory in the background 
of a D-brane and $B_{ij}$ along its world volume described by the action   
\be
S=\frac{1}{4\pi\alpha'}\int_{\Sigma}g_{ij}\del_ax^i\del^ax^j
 -\frac{i}{2}\int_{\del\Sigma}B_{ij}x^i\del_tx^j,
\label{action}
\ee
and showed that in the point-splitting regularization on the world sheet 
the low-energy effective theory is given by the NCYM theory 
with metric $G_{ij}$ and noncommutativity parameter $\theta^{ij}$
given as follows:
\be
\frac{1}{G}=-\frac{\theta}{2\pi\alpha'}+\frac{1}{g+2\pi\alpha' B}.
\label{SW1}
\ee
Moreover, from the fact that in the Pauli-Villars regularization 
it is described by the ordinary YM theory ($\theta=0$), they pointed out 
that both YM theories arise from the same two-dimensional field theory 
regularized in different ways  and are physically equivalent.   
Putting this observation forward, they predicted that for all values 
of $\theta$ there exists an equivalent description of NCYM theory. 
This degrees of freedom is conjectured to be captured by introducing 
a two-form field $\Phi$ as 
\be
\frac{1}{G+2\pi\alpha'\Phi}=-\frac{\theta}{2\pi\alpha'}
                            +\frac{1}{g+2\pi\alpha' B}.
\label{SW2}
\ee
Although setting $\Phi=0$ reproduces (\ref{SW1}), in the general case of 
$\Phi\neq 0$ we get NCYM theory with a different value of $\theta$. 
$\Phi$ corresponds to the degrees of freedom of a magnetic background 
in the NCYM theory and its existence is naturally required from  
the Morita equivalence which is an equivalent relation between different 
NCYM theories\cite{CDS,ho,PS}. In \cite{PS} the degrees of freedom to change 
$\Phi$ and $\theta$ in a suitable way keeping the NCYM theory 
physically invariant are argued to exist. From the point of view of 
the noncommutativities of D-branes, as mentioned in the introduction, 
the noncommutativity $\theta$ of the NCYM theory reflects that of the 
base space coordinate. In the present example the base space is nothing but 
the D-string world volume and in this sense $\theta$ can be considered 
to parametrize the longitudinal noncommutativity. 

On our D-string world volume there is a constant $U(1)$ gauge field strength 
as in (\ref{spiden}) and thereby the second term in (\ref{action}) are 
modified as $B'_{ij}\rightarrow B'_{ij}+\epsilon_{ij}/k$. 
Noting this, we obtain from (\ref{D-1andD1}), (\ref{SW2})
\be
  \frac{1}{G'+2\pi\alpha'(B'+\epsilon/k)} 
= \frac{1}{G+2\pi\alpha'\Phi}+\frac{\theta}{2\pi\alpha'}. 
\label{basiceq}
\ee
This is the equation which relates the moduli of the D-instanton configuration 
$G'$, $B'$, $k$ to the parameters in the NCYM theory $G$, $\theta$, $\Phi$. 
A similar formula was also obtained in \cite{IIKK}. In what follows, we will 
define $\cF'$ as
\be
\cF'\equiv B'+\frac{1}{k}\epsilon.
\label{cF'}
\ee

In some particular cases, $\theta$ and $\Phi$ are expressed in a simple form 
from (\ref{basiceq}): \\
(1) $\Phi=0$. 
\be
\frac{1}{\theta^{12}}=-\cF'_{12}-\frac{\det G'}{(\str)^2\cF'_{12}},
\ee
this corresponds to the NCYM theory obtained in the point-splitting 
regularization in \cite{SW} \\
(2) $\theta=0$.
\be
\Phi=\cF',
\ee
this corresponds to the ordinary YM theory (however, with the magnetic 
background $\Phi$ ) obtained in the Pauli-Villars regularization. \\
(3) $\theta=1/\cF'$.  
\be
\Phi=-\cF',
\ee
this corresponds to the exact solution or the solution in the zero-slope limit 
argued in \cite{SW}
\be
\alpha'\sim\epsilon^\frac{1}{2}\rightarrow 0,~~~~~
g_{ij}\sim\epsilon\rightarrow 0.
\label{SWlimit}
\ee
In fact, in \cite{SW} one of the exact solutions to (\ref{SW2}) is proposed as 
\be
\theta=\frac{1}{B},~~~
G=-(\str)^2B\frac{1}{g}B,~~~
\Phi=-B,
\label{exactsol}
\ee
and it is pointed out that this is also the solution in the limit 
(\ref{SWlimit}). 

In the intermediate region (general case) of interest at present, 
(\ref{basiceq}) leads to a quadratic equation for $\Phi_{12}$ 
and its solution is determined in a unique way from the requirement that 
it should reproduce the above special cases. It reads 
\be
\Phi_{12}= \frac{1}{\theta^{12}}
          -\frac{  \cF'_{12}+1/\theta^{12}  }
                { (\cF'_{12}\theta^{12}+1)^2
                  +\det G'\left(\theta^{12}/\str\right)^2   }.
\label{generalsol}
\ee
This is the most general exact solution to (\ref{basiceq}) in our case. 
For given $\cF'$, if we choose $\theta$ in an arbitrary fixed value, 
we can determine $\Phi$ according to this equation. 
In the low-energy limit (zero-slope limit), as we see above, 
$\theta^{12}=-1/(B'_{12}+1/k)$ and in this sense the transverse 
noncommutativity $k$ and the momentum noncommutativity $B'_{12}$ 
of the D-instantons are encoded into the longitudinal noncommutativity 
$\theta$ of the D-string in a special combination. However, in general cases, 
the $\Phi$ degrees of freedom make their relation complicated as in 
(\ref{generalsol}). Note that in such cases the situation that 
transverse and momentum noncommutativity are combined into 
the longitudinal one only in the special combination is the same. 
This fact seems to imply that in defining the nonperturbative string theory 
in terms of the D-instanton degrees of freedom, it is necessary to introduce 
not only the transverse noncommutativity as has been formulated so far, 
but also the momentum noncommutativity on an equal footing.

\section{The General Coordinate Transformation on the D-string World Volume 
and the $\Phi$ Degrees of Freedom}

In this section from the point of view of the D-string world volume theory 
we reconsider the degrees of freedom of $\Phi$ conjectured in \cite{SW} 
which changes the value of $\theta$ keeping NCYM theory intact. 
As discussed in \cite{ishi,ishi2,cornalba}, it is known that 
if we choose the coordinate on the world volume in such a way 
that the $U(1)$ field strength $\cF_{ij}$ is given as $\epsilon_{ij}/k$ 
or $B_{ij}$ by using the coordinate transformation, 
the resulting coordinate shows the longitudinal noncommutativity 
in the form of (\ref{qplane}) with $\theta^{ij}=-k\epsilon^{ij}$ 
or $\theta^{ij}=(B^{-1})^{ij}$, respectively. 
In the following we generalize the argument in \cite{cornalba}. 
Namely, by making the general coordinate transformation on the D-string 
world volume considered in the previous section, we change the parametrization 
of the world volume from that in the static gauge described by 
(\ref{D-1andD1}) to that in a gauge in which $\cF_{12}$ is equal to 
a certain constant value 
and reconsider the world volume theory in this gauge. Then 
it is found that if it is possible to fix the value of $\cF_{12}$ arbitrarily 
not restricted to $1/k$ or $B_{12}$, we get a description 
with an arbitrary longitudinal noncommutativity of the same world volume 
theory.  

Let us denote the D-string world volume coordinate in the static gauge 
considered in section 2 as $x^i$:
\be
X^i(x)=x^i,
\label{staticgauge}
\ee
where $X^i$ is the embedding function into the target space. 
In the static gauge a dynamical field on the world volume is a $U(1)$ 
gauge field $a_i(x)$. Combining the result in (\ref{D-1andD1}), 
the total field strength is given by  
\be
\cF_{ij}(x)=\cF'_{ij}+f_{ij}(x)
           =\cF'_{ij}+\del_{x^i}a_j(x)-\del_{x^j}a_i(x),
\ee
where $\cF'$ is a constant field strength defined in (\ref{cF'}). 

To begin with, we consider the case in which $a_i(x)=0$. 
For an arbitrary non-zero constant antisymmetric tensor field 
$Y_{ij}=Y\epsilon_{ij}$ on the D-string world volume, 
we make a coordinate transformation on the world volume 
and choose its new coordinate $\si^i$ under which 
a new field strength $\tcF_{ij}(\si)$ is given by $Y_{ij}$: 
\be
Y_{ij}=\tcF_{ij}(\si)=J_i^kJ_j^l\cF'_{kl}(x(\si)),
\label{newcoord}
\ee
where $J_i^k$ is the Jacobian matrix associated with our coordinate 
transformation:
\be
J_i^k=\jac{x^k}{\si^i}.
\ee
Eq. (\ref{newcoord}) means that  
\be
Y=J\cF'J^T,
\label{sigmacond}
\ee
and this is the condition which the coordinate $\si$, in other words, 
the coordinate transformation $x\rightarrow \si$, should satisfy. 
Below we will assume that for an arbitrary $Y_{ij}$ we can choose such a 
coordinate $\si^i$. The special case $B=0$ and $Y=1/k$ is considered in 
\cite{ishi,ishi2} and the case $k\rightarrow\infty$ and $Y=B_{12}$ 
corresponds to \cite{cornalba}. 

Next when there is a small fluctuation of $U(1)$ gauge field $a_i(x)$ in 
the static gauge (\ref{staticgauge}), what is the counterpart of 
its degrees of freedom in the $\si$ coordinate ($\tcF=Y$ gauge)? 
In the presence of the $a_i(x)$, we denote a new coordinate $\si'^i$
which satisfies the gauge condition 
\be
Y_{ij}=\tcF_{ij}(\si')=J_i^kJ_j^l\cF_{kl}(x(\si')),
\label{dycoord}
\ee
as 
\be
\si'^i=\si^i-d^i(\si).
\label{defofd}
\ee
Here $\si^i$ is the coordinate which satisfies (\ref{newcoord}). 
$d^i(\si)$ defined here is a dynamical field corresponding to $a_i(x)$ 
in this gauge. Define 
\be
{J'}_i^k=\jac{x^k}{\si'^i}=J_i^k+\dJ_i^k,
\ee
then (\ref{dycoord}) can be rewritten as 
\bea
Y_{ij} & = & (J+\dJ)_i^k(J+\dJ)_j^l(\cF'_{kl}+f_{kl}) \nono \\
       & = & Y_{ij}+\dJ_i^kJ_j^l\cF'_{kl} 
                   +J_i^k\dJ_j^l\cF'_{kl}+J_i^kJ_j^lf_{kl},
\eea  
here we have assumed that $\dJ\sim O(a)$ and kept terms up to the first 
order in $a$. 
This equation leads to 
\be
JfJ^T=-\dJ\cF' J^T-J\cF'\dJ^T.
\label{jfjt}
\ee
We note that 
\be
{J'}_i^j=J_i^k+(\del_id^k(\si))J_k^j,~~~\del_i\equiv\jac{~}{\si^i},
\ee
and thus if we define $M_i^k=\del_id^k(\si)$, 
then $J'=(1+M)J$, namely, $\dJ=MJ$. Substituting this into (\ref{jfjt}) 
yields 
\be
JfJ^T=-MY+(MY)^T, 
\ee
here we have used (\ref{sigmacond}). 
{}For components, this equation gives  
\be
f_{12}=-Y_{12}\det J^{-1}\tr M=-Y_{12}\det J^{-1}\del_id^i.
\ee
Using $\det J=Y_{12}/\cF'_{12}$ which can be easily obtained by 
(\ref{dycoord}), we find 
\be
-\cF'_{12}\del_id^i=f_{12}=\del_{x^1}a_2-\del_{x^2}a_1.
\ee
Noting that $\del_{x^i}=(J^{-1})_i^j\del_j$, this equation gives 
\be
Y_{12}\del_id^i=-\epsilon_{ij}\del_i(J_j^ka_k).
\ee
Thus we get 
\be
d^i=(\Yi)^{ij}\del_jx^ka_k=(\Yi)^{ij}\tilde{a}_j,
\label{NCYMfield1}
\ee
where we have introduced
\be
\tilde{a}_j(\si)=\jac{x^k}{\si^j}a_k(x(\si)),
\ee 
and ignored a difference of the $U(1)$ gauge degrees of freedom 
\be
d^i\rightarrow d^i+(\Yi)^{ij}\del_j\lambda.
\ee
This result certainly reproduces those obtained in 
\cite{ishi,ishi2,cornalba} as special cases and is a natural 
extension of theirs. 

Now let us derive the action and its symmetry in the gauge (\ref{newcoord}). 
Following the argument in \cite{cornalba}, 
we begin with the Born-Infeld action for our D-string in this gauge
\be
S=T\dDst\sqrt{\det(\tg_{ij}+\str\tcF_{ij})},
\ee
where $\tg_{ij}$ is the induced metric for the embedding function 
$X^i(\si)=x^i(\si)$ 
\be
\tg_{ij}=\del_ix^a\del_jx^bg_{ab}. 
\ee
Using the gauge condition (\ref{newcoord}), this action becomes
\bea
S & = & \mbox{const.}+\frac{T}{4}\sqrt{(\str)^2Y_{12}^2}
                       \dDst\frac{1}{(\str)^2}(\Yi)^{il}(\Yi)^{jk}
                       \del_ix^a\del_jx^b\del_kx^c\del_lx^dg_{ab}g_{cd}
                     +\cdots 
                                                             \nono   \\
  & = & \mbox{const.}-\frac{T}{4}\sqrt{(\str)^2Y_{12}^2}
                       \dDst\frac{1}{(\str)^2}\PB{x^a}{x^d}\PB{x^b}{x^c}
                       g_{ab}g_{cd}
                     +\cdots,
\eea
where we have introduced the Poisson bracket associated with $Y^{-1}$ as
\be
\PB{A}{B}=i(\Yi)^{ij}\del_iA\del_jB.
\label{PB}
\ee 
In particular, 
\be
\PB{\si^i}{\si^j}=i(\Yi)^{ij}.
\label{ncstr}
\ee
In the presence of the small fluctuation $d^i(\si)$ defined in (\ref{defofd}), 
we can see from (\ref{NCYMfield1}) that the embedding coordinate 
becomes 
\be
x^a\rightarrow x^a+\delta x^a=x^a+J_i^a(\Yi)^{ij}\ta_j.
\label{deformation}
\ee
Then it is easy to check that 
\be
-i\PB{x^a}{x^d}\rightarrow (J^T\Yi J)^{ad}-(J^T\Yi)^{ai}\tf_{ij}(\Yi J)^{jd},
\ee
where 
\be
\tf_{ij}=\del_i \ta_j-\del_j\ta_i-i\PB{\ta_i}{\ta_j}.
\ee
Thus we obtain an action for the fluctuation $\ta_i$ as
\be
\frac{T}{4}\sqrt{\frac{Y_{12}^2}{(\str)^2}}
\dDst(J^T\Yi)^{aj}(\tf_{jl}-Y_{jl})(\Yi J)^{ld}
     (J^T\Yi)^{bk}(\tf_{ki}-Y_{ki})(\Yi J)^{ic}g_{ab}g_{cd}.
\ee
Define 
\bea
G^{ij} & = & -\frac{1}{(\str)^2}(\Yi J)^{ia}g_{ab}(J^T\Yi)^{bj} 
            =-\frac{1}{(\str)^2}(\Yi\tg\Yi)^{ij},
\label{openstringmetric} \\
G_{s} & = & g_s \det(2\pi\alpha'Y\tg^{-1})^{\frac{1}{2}},
\label{openstringcouping}
\eea
then the action takes the form of 
\bea
\lefteqn{\frac{1}{g_{YM}^2}\dDst\sqrt{\det G_{ij}}G^{ij}G^{kl}\frac{1}{4}
                           (\tf_{ik}-Y_{ik})(\tf_{jl}-Y_{jl})} & & 
\nono \\
 & &   = \frac{1}{g_{YM}^2}\dDst\sqrt{\det G_{ij}}G^{ij}G^{kl}\frac{1}{4}
                            \tf_{ik}\tf_{jl}+\mbox{total derivative},
\label{YMaction}
\eea
where 
\be
g_{YM}^2=\frac{\str}{G_s}.
\ee
This is the action of the NCYM theory with $\theta=\Yi$ up to the second 
order in $\Yi$. We believe that generalization of the arguments given 
in \cite{oku} would work in our case and it provides the exact NCYM theory 
with $\theta=\Yi$ in which the Poisson bracket (\ref{PB}) is replaced 
by the Moyal bracket. 
Thus we have verified that at least for small fluctuations 
the D-string world volume theory in the gauge (\ref{dycoord}) is 
described by the NCYM theory with noncommutativity $\Yi$ 
and that $\ta_i$ corresponds to the NC gauge field. 
It is worth noticing that (\ref{openstringmetric}) and 
(\ref{openstringcouping}) are natural generalization of 
the open string metric and open string coupling constant 
in the zero-slope limit given in \cite{SW}. 

As another evidence that we have NCYM theory with $\theta=\Yi$,            
let us analyze the symmetry of (\ref{YMaction}). It is argued 
in \cite{ishi,ishi2,cornalba} that the gauge condition of the form 
(\ref{newcoord}) does not completely fix the whole reparametrization 
invariance and there is a residual one, which can be interpreted 
as the noncommutative gauge invariance (\ref{nctransf}) in terms of 
$\ta_i$ defined like (\ref{deformation}). In the present case, we also have 
residual diffeomorphism invariance after fixing the gauge as in 
(\ref{newcoord}). Namely, if we make the coordinate transformation 
$\si^i\rightarrow \si^i+V^i(\si)$ using a vector field $V^i(\si)$ 
on the world volume, the condition that this coordinate preserves 
the gauge condition is 
\be
Y_{kj}\del_iV^k(\si)+Y_{ik}\del_jV^k(\si)=0,
\ee
which implies that there exists a scalar field $\rho(\si)$ on the world 
volume such that 
\be
V^i=(\Yi)^{ij}\del_j\rho.
\ee
Then $x^i(\si)$ is transformed as 
\be
x^i\rightarrow x^i-(\Yi)^{jk}\del_j\rho\del_kx^i=x^i+i\PB{\rho}{x^i}. 
\label{residiff}
\ee
This and (\ref{ncstr}) strongly suggest that the world volume coordinate 
in the gauge (\ref{newcoord}) has the longitudinal noncommutativity 
$\theta=\Yi$ in the form of (\ref{qplane}). 
If there exists the fluctuation $\ta_i(\si)$, $x^i$ is modified according to 
(\ref{deformation}) and the transformation (\ref{residiff}) becomes 
\be
x'^i\rightarrow x'^i+i\PB{\rho}{x'^i}\equiv x'^i+\delta x'^i.
\label{residiff'}
\ee
In terms of $\ta_i(\si)$, (\ref{residiff'}) can be rewritten as 
\be
\delta\ta_i=\del_i\rho+i\PB{\rho}{\ta_i},
\ee
where we have used $\delta x'^i=J_j^i(\Yi)^{jk}\delta\ta_k$ 
and $\del_i\del_j x^k=0$, because it is possible to 
choose $J$ as a constant matrix in the case of a constant $\cF'$. 

Thus we conclude that for an arbitrary non-zero constant antisymmetric 
field $Y_{ij}$, if we choose a coordinate $\si$ such that $\tcF(\si)=Y$ 
by means of the general coordinate transformation on the world volume, 
the world volume theory in this coordinate is described by the NCYM 
theory with $\theta=\Yi$ and correspondingly, the world volume has 
the longitudinal noncommutativity $(\Yi)^{ij}$.  

We have seen that the longitudinal noncommutativity $\theta$ can be 
arbitrarily varied by the way of fixing the diffeomorphism invariance 
on the world volume. As emphasized in the previous section, 
from the point of view of NCYM theory, the magnetic background $\Phi$ 
is responsible for the degrees of freedom to change the value of $\theta$. 
Therefore, it is conjectured that the degrees of freedom of $\Phi$ 
should correspond to those of diffeomorphism on the world volume 
which is noncommutative in general. In order to see that this is the case, 
let us briefly discuss the derivation of (\ref{generalsol}). 
The following argument is a generalization of that 
given in \cite{IIKK}. It is natural to guess that the D-string configuration 
in the $\tF'=Y$ gauge is equivalent to that in the static gauge 
in the background metric $g_{ij}=G'_{ij}$, antisymmetric tensor field 
$\tF'-Y$, and $U(1)$ field strength $Y$. The T-duality transformation 
in 1,2-directions maps this to a noncommutative D-instanton configuration
\be
[\bq^1, \bq^2]=-i(\Yi)^{12},
\label{Tdualconfig}
\ee
in the T-dual background
\be
\bG'^{ij}+\str\bB^{ij}=\frac{1}{G'+\str(\tF'-Y)}.
\ee
Similarly to above and \cite{IIKK}, for fluctuations $\delta\bq^i$ 
around (\ref{Tdualconfig}), we define $\ba_i$ as
\be
\delta\bq^i=(\Yi)^{ij}\ba_j(\bq).
\ee
Substituting this in the original effective action of D-instantons
for slowly varying fields\footnote
{We adopt a naive generalization of the abelian Born-Infeld action.} 
\be
S\sim\Tr\sqrt{\det(\bG^{ij}+\str\bB^{ij}+i(\str)^{-1}[\bq^i,\bq^j])},
\ee
and expressing the functions of $\bq^i$ as c-number functions 
whose product is given by the $*$-product in (\ref{*product}) 
with $\theta=\Yi$, we get the noncommutative Born-Infled action 
in the following form:
\be
S_{\Phi}\sim\sqrt{\det(\hat{G}+\str (f+\Phi))},
\ee
where 
\be
\Phi=-Y-(\str)^2Y\bB Y,
\ee
which reproduces (\ref{generalsol}) with $\theta=\Yi$. 

Now let us summarize above arguments using boundary states 
in accordance with \cite{ishi,ishi2}. 
As constructed in \cite{ishi}, a boundary state corresponding to (\ref{wvnc}) 
is given by 
\be
\ket{B}_{-1}=\tr P\exp\left(-i\int_0^{2\pi}ds P_i(s)q^i
                      \right)\ket{X^i=0}_{-1},
\label{D-1bs}
\ee 
where $P_i$ is the canonical momentum of string and 
$\ket{X^i=0}_{-1}$ is the Dirichlet boundary state:
\be
X^i(s)\ket{X^i=0}_{-1}=0.
\ee
In the path integral representation, (\ref{D-1bs}) can be rewritten as 
\be
\ket{B}_{-1} = \int [dq^1dq^2]\exp\left(\frac{i}{k}
               \int_0^{2\pi}ds q^1(s)\partial_sq^2(s)
             -i\int_0^{2\pi}ds P_i(s)q^i(s)
                                 \right)\ket{X^i=0}_{-1}.
\ee
Generalizing this, a general configuration of D-instantons 
\be
X^{\mu}=\phi^{\mu}(q)~~~(\mu\geq 1),
\ee
should correspond to a boundary state 
\be
\ket{B}_{-1} = \int [dq^1dq^2]\exp\left(\frac{i}{k}
               \int_0^{2\pi}ds q^1(s)\partial_sq^2(s)
             -i\int_0^{2\pi}ds P_{\mu}(s)\phi^{\mu}(q^i(s))
                                  \right)\ket{X^i=0}_{-1}.
\label{generalD-1bs}
\ee
On the other hand, in the D-string picture, the world volume theory 
consists of a $U(1)$ gauge field $A_i$ on the world volume 
and scalar fields $\phi^a$ ($a\geq 3$) which describe the transverse 
coordinates of the world volume. In the static gauge $\phi^i(q)=q^i$, 
the boundary state corresponding to this general configuration is given by 
\bea
\ket{B}_{1} & = & \int [dq^1dq^2]\exp
                  \left(i\int_0^{2\pi}ds A_i(q(s))\partial_sq^i(s)
                -i\int_0^{2\pi}ds (P_iq^i(s)+P_a(s)\phi^{a}(q(s)))
                  \right)  \nono \\
            &   & \times  \ket{X^i=0}_{-1},
\label{generalD1bs}
\eea
which coincides with (\ref{generalD-1bs}) under an identification 
$F_{12}=1/k$. Thus we see that in both boundary state $q^i$ plays a role of 
the parametrization of the world volume. Now following \cite{ishi2}, 
let us consider the most general boundary state which includes all of the 
fields $A_i$, $\phi^{\mu}$ which appeared in (\ref{generalD-1bs}), 
(\ref{generalD1bs})
\be
\ket{B} = \int [dq^1dq^2]\exp
              \left(i\int_0^{2\pi}ds A_i(q(s))\partial_sq^i(s)
                   -i\int_0^{2\pi}ds P_{\mu}(s)\phi^{\mu}(q(s))
              \right)\ket{X^i=0}_{-1}.
\label{generalbs}
\ee
Then it is easy to check that this boundary state is invariant under 
the reparametrization of the world volume coordinate $q^i$\cite{ishi2}. 
Moreover, by using this invariance, when we fix $q^i$ in the static gauge 
$\phi^i(q)=q^i$, (\ref{generalbs}) reproduces (\ref{generalD1bs}), while 
if we fix $q^i$ in such a way that $F_{12}=1/k$, it reproduces 
(\ref{generalD-1bs}). In the latter case, as we have seen in (\ref{ncstr}), 
$q^i$ become noncommutative coordinates. In this sense, an original 
configuration described by (\ref{generalbs}) has the invariance 
of the diffeomorphism on the (noncommutative, in general) world volume
\cite{ishi2}. According to the notation in \cite{ishi}, we denote the group 
which consists of such diffeomorphisms as $Diff$. $Diff$ has a subgroup 
of diffeomorphisms which preserve the value of $\cF$. We denote it as 
$Diff_{\cF}$. This is exactly the symmetry in the D-instanton picture 
and is inherited from the original $U(\infty)$ symmetry\cite{ishi2} 
\be
\delta q^i=i[\epsilon,q^i]. 
\ee
As we showed above, $Diff_{\cF}$ can be also interpreted as 
the noncommutative gauge symmetry\cite{ishi,ishi2,cornalba}. 
Arguments in this section implies that the residual symmetry of 
the reparametrization invariance which the original system (\ref{generalbs}) 
has, namely, 
\be
\frac{Diff}{Diff_\cF},
\ee
corresponds to a hidden `symmetry' which is responsible for changing 
the longitudinal noncommutativity and is conjectured to be equivalent to 
the degrees of freedom of $\Phi$.

\section{Discussions}

Let us discuss the meaning of our results from the point of 
view of searching a nonperturbative mechanism and definition 
of string theory. 

We have shown that the `symmetry' to change the longitudinal noncommutativity 
results from the degrees of freedom of the way of choosing parametrizations 
on a noncommutative world volume. Thus starting with a configuration 
like (\ref{generalbs}) gives a clue to the proof of 
(\ref{SW2}) or (\ref{basiceq}). In fact, there have been already some results 
in this approach\cite{oku}. 
Moreover, it is expected to clarify relations or symmetries 
between various noncommutativities.  In particular, although we concentrated 
on the relation between the transverse noncommutativity and the longitudinal 
one in the above discussion using boundary states, 
it would be interesting to construct a boundary state which also includes 
the momentum noncommutativity.\footnote{
{}For example, in \cite{KK} such a boundary state is concretely constructed.}
If it is possible, it would make clear interrelations 
between the noncommutativities and serve for the nonperturbative formulation 
of string theory. 

On the other hand, from the point of view of D-brane matrix models, 
the noncommutativities may provide natural regularizations. 
If we see the lower-dimensional D-brane matrix models 
as candidates for the nonperturbative definition of string theory 
like\cite{BFSS,IKKT,Mald}, these must be well-defined field theories. 
It would be interesting to examine a possibility 
that various noncommutativities associated with D-branes provide 
natural regularizations for them. In this sense, it would be 
important to define a field theory on a noncommutative space 
and to clarify a role played by the noncommutativity as a regularization
\cite{master}. 

In relation to the work of \cite{SW}, we believe that the gauge choice 
(\ref{newcoord}) would correspond to a certain regularization of 
the two-dimensional field theory which interpolates between 
the Pauli-Villars and the point-splitting, for example, a hybrid point 
splitting proposed in \cite{andorn}. We hope to return to this issue 
in future works.

\section*{Acknowledgements}

The author would like to thank O. Andreev, M. Kato and N. Ishibashi 
for valuable discussions and useful comments. This work was also influenced 
so much by the Summer Institute '99 at Yamanashi and the workshop at YITP, 
Japan. The author is grateful to the organizers of these workshops 
for providing such opportunities.

\end{document}